\documentstyle[12pt]{article}
\newcommand{\be}{\begin{equation}}
\newcommand{\ee}{\end{equation}}
\newcommand{\ba}{\begin{eqnarray}}
\newcommand{\ea}{\end{eqnarray}}
\renewcommand{\thefootnote}{\fnsymbol{footnote}}
\begin{document}
\begin{center}
{\Large 
Exactly Solvable Three-body 
SUSY Systems with Internal Degrees of Freedom}\\
\vspace{1cm}
{\large F. Cannata} $^{a,}$\footnote{{\it E-mail:} cannata@bo.infn.it},
{\large M. Ioffe} $^{b,}$\footnote{{\it E-mail:} ioffe@snoopy.phys.spbu.ru}\\ 
$^a${\small\it Dipartimento di Fisica and INFN, Via Irnerio 46,
40126 Bologna, Italy}\\
$^b${\small\it Department of Theoretical
Physics, Institute of Physics, University of Sankt-Petersburg,
Ulyanovskaya 1, Sankt-Petersburg 198904, Russia}
\end{center}
\vspace{1cm}
\noindent 
{\bf Abstract}

\bigskip

{\small 
The approach of multi-dimensional SUSY Quantum Mechanics is used 
in an explicit construction of exactly solvable $3$-body 
( and quasi-exactly-solvable $N$-body ) matrix problems on a line.
From intertwining relations with time-dependent operators,
we build exactly solvable non-stationary scalar and $2\times 2$ matrix 
$3$-body models which are time-dependent extensions of the
Calogero model. Finally, we investigate the invariant operators associated
to these systems. 
} 
\bigskip
\setcounter{footnote}{0}
\renewcommand{\thefootnote}{\arabic{footnote}}
\section*{\large\bf 1.\quad Introduction}
\hspace*{2ex} 

During the last three decades exactly solvable $N$-body problems
have provided useful tools to investigate formal algebraic properties 
with applications to different branches of Physics.
The most widely studied model is the so called Calogero model 
\cite{calogero} and its various
generalizations which essentially are many-body extensions of the 
one-dimensional singular harmomic oscillator model.
Calogero-like models have been developed incorporating 
different root systems \cite{sasaki}, $q$-deformations \cite{q},
{\cal PT}-symmetric generalizations \cite{znojil}, many-body forces
\cite{applications}, multi-dimensions \cite{multi} and internal 
degrees of freedom with potentials which couple them (matrix potentials)
\cite{matrix}. Even if coupled channel problems in general have a 
continuum \cite{ci}
and discrete spectrum, in the Calogero-like models (with harmonic attraction)
the spectrum is purely discrete reflecting an essentially confining
dynamics. Physical applications of this dynamics have been elaborated
in the context of localized systems like, in particular, Paul traps and
quantum dots \cite{dots}.

An important generalization of the Calogero model is 
its supersymmetrization \cite{freedman}. 
Supersymmetric Quantum Mechanics (SUSY QM) \cite{reviews} 
is a suitable framework 
to discover and investigate one-dimensional, one-particle exactly-solvable
models \cite{infeld}. The same strategy applies to 
systems with multiple degrees of freedom. 
Multi-dimensional SUSY QM, first constructed in \cite{abi}, 
leads to a superHamiltonian which includes a chain of matrix Hamiltonians
(cf. coupled channels or internal degrees of freedom, like spin \cite{ai}). 
Supersymmetry ensures that the spectral properties and the 
eigenfunctions of the Hamiltonians belonging to the chain are algebraically 
interrelated. This analysis can be reinterpreted with reference to a
one-dimensional multi-particle problem enlarging classes of many-body
exactly solvable problems in a similar way as for the one-dimensional 
one-particle problems \cite{am}.

We start from the superpotential of the Calogero system which corresponds 
to two exactly-solvable scalar components of the superHamiltonian.
They are intertwined to the neighbouring matrix components 
of the superHamiltonian.
This implies that a part of the spectrum for both matrix potentials
and the corresponding wave functions are known. Thus from solvable
scalar models by supersymmetric techniques 
quasi-exactly-solvable matrix problems are generated.
This approach generates matrix $N$-particle models which can 
be considered in the context  
of recently constructed scalar quasi-exactly-solvable
\cite{shifman} and so called partially solvable \cite{calogeronew} models. 

In Section 2 we review the basic aspects of multi-dimensional 
SUSY QM \cite{abi} and introduce its reinterpretation in terms of  
multi-particle one-dimensional SUSY QM (see details in the recent paper
\cite{am}).
In particular, we focus attention on models with exactly-solvable
scalar components of the superHamiltonian. 

In Section 3 starting from
the Calogero model \cite{calogero} we analyze its SUSY extension which
includes quasi-exactly-solvable \cite{turbiner} 
$ N-$particle matrix models.
Furthermore,  we study 3-body problems in detail 
because the properties of the chain
of the components of the superHamiltonian simplify considerably so that
the spectrum and the wave functions for the (only) matrix Hamiltonian 
are fully determined from those of (two) scalar Hamiltonians. 

In addition to exactly-solvable stationary problems we consider 
also time-dependent potentials and, correspondingly, exactly-solvable
time-dependent problems. In the context of one-dimensional one-particle 
SUSY QM (and Darboux transformation) such problems were investigated 
in \cite{tdse} .  
In non-stationary SUSY QM  supercharges (of first and second order
in space derivatives) commute
with the non-stationary Schr\"odinger superoperator and 
intertwine consecutive
components of the supersymmetric chain. 
Following methods developed in recent investigations of the time-dependent
harmonic oscillator model and its generalizations \cite{china},
in Section 4 we construct time-dependent $3$-particle 
solvable problems. In this Section we achieve our main goal after
having prepared in the previous Sections the relevant framework. 
These results
can be interpreted as time-dependent generalization of the SUSY Calogero model, 
which can be shown to be solvable
by introducing unitary intertwining operators (non-polynomial in 
derivatives). An extension of this method to the $N$-body Calogero model
described in Section 2 leads to time-dependent quasi-exactly-solvable
matrix models.
While the Calogero model is a many-body generalization of the 
"singular" harmonic oscillator model, its time-dependent version, which
we study, are correspondingly extensions of the
oscillator problem with time-dependent parameters. 
This last problem has attracted 
much interest in the literature \cite{china}, \cite{nieto} and
has applications in different areas of Physics.

\section*{\large\bf 2.\quad Multi-dimensional SUSY QM and $N$-particle
quasi-exactly-solvable stationary problems}
\hspace*{2ex}
The supersymmetric quantum system for arbitrary number of dimensions
$N$ consists \cite{abi}  of the superHamiltonian $H_S$ and the 
supercharges $Q^{\pm}$
with the algebra (SUSY QM algebra):
\ba
H_S = \{ Q^+,Q^-  \} ,\label{tri} \\
(Q^+)^2=(Q^-)^2=0, \label{che}
\ea
\be
[H_S,Q^\pm]=0. \label{sus}
\ee
An explicit realization is given 
by\footnote{Here and below the indices $i,j,k,\ldots $ range from 1 to $N $.}:
\ba
H_S = \frac{1}{2}(-\Delta + \sum_{i=1}^N(\partial_i W)^2-\Delta W)
+\sum_{i,j=1}^N\psi_i^+\psi_j\partial_i\partial_j W; \qquad \Delta\equiv
\sum_{i=1}^N \partial_i\partial_i;\quad \partial_i\equiv \partial / 
\partial x_i;
\label{HSS}
\ea
\ba
Q^{\pm}\equiv
 \frac{1}{\sqrt{2}}\sum_{j=1}^N \psi_{j}^\pm (\pm\partial_j+\partial_j W), 
\label{lkq}
\ea
where $\psi_i,\ \psi_i^+$ are standard fermionic operators:
\be
\{ \psi_i,\psi_j\} =0,\qquad
\{ \psi_i^+,\psi_j^+\} =0, \qquad
\{ \psi_i,\psi_j^+\} =\delta_{ij}.
\nonumber
%\label{ant}
\ee
The dynamics of a particular SUSY QM system is determined by a 
superpotential
$ W$, which depends on $N$ coordinates $(x_1,\ldots,x_N)$.

In general, solvable scalar models in  multi-dimensional 
Quantum Mechanics for one particle admit 
simple separation of 
variables and are, therefore, reducible to one-dimensional problems. 
For this reason from now on we will alternatively  
interpret multi-dimensional SUSY QM as a multi-particle 
problem on a line because one knows classes of solvable models
(Calogero model, Sutherland model and others \cite{sasaki}) 
which do not admit such a straightforward separation.

For $N$-particle systems on a line it is natural to consider \cite{am}
superpotentials with a separable centre-of-mass motion (CMM),
satisfying the condition:
\be
 W(x_1,...,x_N)=w(x_1,\ldots,x_N)+W_0\biggl(x_1+...+x_N\biggr); \quad
\sum_{j=1}^{N}\partial_j w(x_1,\ldots,x_N)=0, \label{sep}
\ee
i.e. the first term $ w(x_1,\ldots,x_N)$ does not depend on 
$ \sum^{N}_{i=1}x_{i}. $
We will restrict ourselves to 
superpotentials (\ref{sep}) with 
$ W_{0}=0,$ or, equivalently, 
$ \sum^{N}_{k=1}\partial_{k}W(x_{1},...,x_{N})=0. $

For the superpotentials (\ref{sep}) one can use 
the well-known Jacobi
coordinates\footnote{From this moment on, the variables denoted 
by letters $a,b,c,\ldots$ range from 1 to $(N-1)$.} \cite{reed}:
\ba
&y_b&=\frac{1}{\sqrt{b(b+1)}}(x_1+\ldots+x_b-bx_{b+1});
\label{bjc} \\
&y_N& = \frac{1}{\sqrt{N}}\sum_{i=1}^N x_i, \nonumber
\ea
or, shortly, $y_k=\sum_{l=1}^NR_{kl}x_l$, where the 
orthogonal matrix $R$ is determined by (\ref{bjc}).
For the supersymmetric systems we introduce also the
fermionic analogues\footnote{The use of these variables has been 
instrumental\cite{am} to clarify the role of the Permutation
Group $S_N$ in SUSY QM. } of the Jacobi variables:
\ba
&\phi_b&=\frac{1}{\sqrt{b(b+1)}}(\psi_1+\ldots+\psi_b-b\psi_{b+1});
\nonumber\\
&\phi_N& = \frac{1}{\sqrt{N}}\sum_{i=1}^N \psi_i, \nonumber
\ea
which satisfy the canonical anticommutation relations:
\be
\{\phi_k,\phi_l\}=0,\qquad
\{\phi_k^+,\phi_l^+\}=0, \qquad
\{\phi_k,\phi_l^+\}=\delta_{kl}.
\nonumber
%\label{fjc}
\ee

In terms of the Jacobi variables the supercharges (\ref{lkq}) 
can be rewritten as:
\ba
Q^{\pm}= q^{\pm} \pm \frac{1}{\sqrt{2}}\phi^{\pm}_{N}
\frac{\partial}{\partial y_{N}}; \nonumber\\
q^{\pm} \equiv \frac{1}{\sqrt{2}}\sum_{b=1}^{N-1}\phi^\pm_b
\biggl(\pm\frac{\partial}{\partial y_b} + 
\frac{\partial}{\partial y_b}w\biggr). \nonumber
\ea

Because
\be
\{ q^{\pm}, \phi^{\mp}_{N}\frac{\partial}{\partial y_{N}} \}=0, 
\nonumber
%\label{ind}
\ee
the free motion of center-of-mass in the superHamiltonian 
can be separated:
\ba
H_S= \{Q^+,Q^-  \} \equiv h - \frac{1}{2} 
\frac{\partial^{2}}{\partial y_{N}^{2}} , \label{HS1}
\ea
where
\ba
  h \equiv \{ q^{+}, q^{-} \} = \frac{1}{2}\sum_{b=1}^{N-1}\biggl(
-\frac{\partial^2}{\partial y_b^2}+
\biggl(\frac{\partial w}{\partial y_b}\biggr)^2-
\frac{\partial^2 w}{\partial y_b^2}\biggr)+
\sum_{b,c=1}^{N-1}\phi_b^+\phi_c
\frac {\partial^
2 w}{\partial y_b \partial y_c},\label{ham}
\ea
is  $(N-1)$-dimensional superHamiltonian expressed in Jacobi variables 
$ y_{1},...,y_{N-1}.$ In the following we will consider
only this reduced superHamiltonian $ h.$

The operator $h$ acting in the fermionic Fock space
\be
\phi_{b_1}^+\ldots\phi_{b_M}^+ |0>;\qquad M<N;\quad b_i<b_j\quad \mbox{for}\ 
i<j ,
\label{ste}
\ee
generated by fermionic creation operators $ \phi_b^+$, conserves
the cor\-res\-pon\-ding fer\-mionic num\-ber 
\be
[ h, N_F ] = 0 \quad \mbox{with}\,\,
N_{F}\equiv\sum^{N-1}_{b=1}\phi^{+}_{b}\phi_{b} . 
\nonumber
%\label{number}
\ee
Therefore, in the basis (\ref{ste}) it has \cite{abi} a
block-diagonal form:
\ba
 h=\pmatrix{
 h^{(0)} & 0 & \ldots & 0 & 0 \cr
 0 & h^{(1)} & \ldots & 0 & 0  \cr
 \vdots & \vdots & \ddots& \vdots & \vdots \cr
 0 & 0 & \ldots & h^{(N-2)} & 0 \cr
 0 & 0 & 0 \ldots & 0 & h^{(N-1)} \cr}, \label{hamm}
\ea
where the matrix operators
$ h^{(M)}$ of dimension $ C_{N-1}^{M}\times C_{N-1}^{M}$
are the projections of $h$ onto the subspaces with fixed
fermionic number $N_{F}=M$. These components are standard
Schr\"odinger operators with matrix potentials and can be obtained from
(\ref{ham}) by a suitable matrix realization of the fermionic
variables $\phi_b.$

The supercharge $q^+$ increases the 
fermionic number from $M$ to $M+1$ and has the  
under-diagonal structure:
\ba
 q^+=\pmatrix{
 0 & 0 & \ldots & 0 & 0 \cr
 q_{(0,1)}^+ & 0 & \ldots & 0 & 0  \cr
 0 & q_{(1,2)}^+  & \ldots & 0 & 0   \cr
 \vdots & \vdots & \ddots& \vdots & \vdots \cr
 0 & 0 & 0 & q_{(N-2,N-1)}^+  & 0  \cr}. \label{qpm}
\ea
 Similarly, $q^-=(q^{+})^{\dagger}$ is an over-diagonal matrix 
operator with nonzero elements\\
 $q_{(M+1,M)}^-=\biggl(q_{(M,M+1)}^+\biggr)^{\dagger}$.

Superinvariance (\ref{sus}) of the superHamiltonian corresponds, in
components, to the intertwining relations:
\ba
h q^+=q^+h  &\Leftrightarrow&
h^{(M+1)} q_{(M,M+1)}^+= q_{(M,M+1)}^+ h^{(M)} 
\nonumber\\
%\label{su1}\\
q^-h = h q^- &\Leftrightarrow&
q_{(M+1,M)}^- h^{(M+1)}= h^{(M)} q_{(M+1,M)}^- 
\nonumber
%\label{su2}
\ea
These relations lead \cite{abi} to important connections
between spectra and eigenfunctions of "neighbouring" Hamiltonians, with
fermionic numbers differing by $ 1.$  In particular, 
$ q_{(M,M+1)}^+$ maps eigenfunctions of $ h^{(M)}$ onto those of
$h^{(M+1)}$ with the same energy $ E_{K} $:
\be
\Psi_{K}^{M+1}(\vec{y}) = q_{(M,M+1)}^{+}\Psi_{K}^{M}(\vec{y});
\quad h^{(M)}\Psi_{K}^{M}(\vec{y})=E_K \Psi_{K}^{M}(\vec{y}) .
\ee
Analogously, $q_{(M,M-1)}^{-}$  maps eigenfunctions
of $h^{(M-1)}$ onto those of $ h^{(M)} $ with the same value of energy
 (see details in \cite{abi}).

In particular, the spectrum of the matrix $(N-1)\times (N-1)$
Hamiltonian $h^{(1)}_{ik}$ consists of two portions, one of
which coincides with the spectrum of the scalar Hamiltonian 
$h^{(0)} .$ Thus if the scalar problem with $h^{(0)}$ 
is solvable the matrix problem with $h^{(1)}_{ik}$ becomes quasi-exactly
solvable \cite{turbiner}. Similarly, the matrix Hamiltonian $h^{(N-2)}$
is also quasi-exactly solvable provided the last (scalar)
Hamiltonian $h^{(N-1)}$ is exactly-solvable.

\section*{\large\bf 3.\quad Stationary solutions of 3-body problem with 
internal degrees of freedom }
\hspace*{3ex}

As a realization of what we presented in Section 2,
we provide an explicit construction
for the $N$-body Calogero model. Substituting the superpotential
\footnote{This form for $W$ is suggested by the ground state
wave function of the conventional Calogero model (see (\ref{calogeroo}))   
and it is known to be a particular choice among
possible alternatives.} which depends only on first $(N-1)$
bosonic Jacobi coordinates $y_1,y_2,...,y_{N-1}$:
\be
W(x_1,x_2,...,x_N) = \alpha \sum_{i\ne j=1}^N(x_i-x_j)^2
+ \frac{\gamma}{2} \sum_{i\ne j=1}^N
\ln |x_i-x_j| = w(y_1,y_2,...,y_{N-1}).  
\label{superpot}
\ee
into (\ref{HSS}), after some manipulations
we obtain apart from a constant energy shift:
\ba
H_S&=&-\frac{1}{2}\Delta^{(N)} + 4\alpha^2N
\sum_{i\ne j=1}^N(x_i-x_j)^2
+\frac{1}{2}\gamma (\gamma +1)
\sum_{i\ne j=1}^N
\frac{1}{(x_i-x_j)^2} +\nonumber\\ 
&+&\gamma
\sum_{i\ne j=1}^N\psi^{\dagger}_i\psi_j \frac{1}{(x_i-x_j)^2}
-\gamma \sum_{i\ne j=1}^N \psi^{\dagger}_i\psi_i \frac{1}{(x_i-x_j)^2}.
\label{calogero}
\ea
After subtraction of the free center-of-mass motion (\ref{HS1})
one obtains 
\footnote{From now on we will use the notation $\partial_i$ for 
$\partial / \partial y_i.$}
a reduced Hamiltonian $h$ from (\ref{ham}), with the
superpotential $w(y_1,y_2,...,y_{N-1}).$
The expression for scalar $h^{(0)}$ can be derived from the superHamiltonian
(\ref{calogero}) by taking into account that the fermionic terms vanish
in the subspace with $N_F=0 :$
\be
h^{(0)}=-\frac{1}{2}\Delta^{(N-1)}_y + 4\alpha^2N
\sum_{i\ne j=1}^N(x_i-x_j)^2
+\frac{1}{2}\gamma (\gamma +1)
\sum_{i\ne j=1}^N
\frac{1}{(x_i-x_j)^2}. 
\label{calogeroo}
\ee
It corresponds to the well-known
exactly solvable $N$-body Calogero model \cite{calogero},\cite{applications}.
As was discussed in the end of the previous
Section, the matrix Hamiltonian $h^{(1)}_{ik}$ is thus 
quasi-exactly solvable and the associated part of its energy levels
coincides with oscillator-like spectrum of (\ref{calogeroo}). 

The last scalar component $h^{(N-1)}$ of the superHamiltonian
(\ref{calogero}) is obtained by its reduction to the subspace
of (\ref{ste}) with maximal fermionic occupation number $N_F=(N-1).$
Only the last fermionic term in (\ref{calogero}) is effective
and $h^{(N-1)}$ coincides\footnote{Let 
us note that the eigenfunctions of $ h^{(0)}$ and 
$h^{(N-1)}$ are not connected directly by supercharges $q^{\pm}$, 
contrary to the hypothesis of the paper \cite{spector} in the 
context of Calogero-like models. In this connection, it was 
recently found \cite{ghosh} that their eigenfunctions are related
by a Dunkl-like differential operator.}
with $h^{(0)}$ after the $\gamma$ into $(-\gamma )$ replacement
\footnote{Note that the ground state energy of the Calogero model
depends on $\gamma$.}.
It is clear that exactly solvability of $h^{(N-1)}$ leads again to
quasi-exactly solvability of the matrix Hamiltonian $h^{(N-2)} .$

For $N=4$ the chain of (\ref{hamm}) consists of two scalar Calogero
Hamiltonians $h^{(0)},\,h^{(3)}$ and two matrix $3\times 3$
Hamiltonians $h^{(1)}_{ik}$ and $h^{(2)}_{ik} ,$ where for example
\cite{abi}: 
\ba
h^{(1)}_{ik}= -\frac{1}{2}\Delta^{(3)}_y +\frac{1}{2}(\partial_i w)^2 +
\frac{1}{2}\pmatrix{
 (\partial_1^2-\partial_2^2-\partial_3^2)w & 
2\partial_1\partial_2w & 2\partial_1\partial_3w  \cr
2\partial_1\partial_2w & 
(\partial_2^2-\partial_1^2-\partial_3^2)w &
2\partial_2\partial_3w   \cr
2\partial_1\partial_3w& 2\partial_2\partial_3w  & 
(\partial_3^2-\partial_1^2-\partial_2^2)w   \cr}
\label{N4}
\ea
and $h^{(2)}_{ik}$ has a similar structure. The Hamiltonian (\ref{N4})
is intertwined to $h^{(0)}$ by 
$
q^+_{(0,1)}\equiv (A_1^-, \, A_2^-,\, A_3^-),
$
where
$A_i^- = (A_i^+)^{\dagger}
\equiv  \frac{1}{\sqrt{2}} (\partial_i + 
\partial_i w(y_1,y_2,y_3) ).
$
Therefore since $h^{(0)}$ is solvable, $h^{(1)}_{ik}$ is quasi-exactly 
solvable. Similar considerations hold concerning the intertwining of
$h^{(2)}_{ik}$ and $h^{(3)}.$ The "non-quasi-exactly solvable"
portions of $h^{(1)}_{ik}$ and $h^{(2)}_{ik}$ coincide \cite{abi}
because of an additional intertwining between them.

It is clear that when the matrix operator $h^{(1)}$ happens 
to coalesce with $h^{(N-2)},$ the quasi-exactly-solvable matrix 
problem becomes exactly solvable. 
This is the case for $N=3$ Calogero model.
We now consider the standard Calogero Hamiltonian for three particles on a 
line with repulsive singular terms.
In terms of Jacobi coordinates
\ba
  y_1=\frac{x_1-x_2}{\sqrt{2}}; \quad
  y_2=\frac{x_1+x_2-2x_3}{\sqrt{6}}  
\nonumber
%\label{jacobi}
\ea
the superpotential $w(y_1,y_2)$ up to an irrelevant constant has the form:
\be
w(y_1,y_2)= 6\alpha (y_1^2+y_2^2) + \gamma 
\ln |y_1(\frac{1}{2}y_1+\frac{\sqrt{3}}{2}y_2)
(-\frac{1}{2}y_1+\frac{\sqrt{3}}{2}y_2)|
\nonumber
%\label{w}
\ee

The Hamiltonian (\ref{calogeroo}) can be rewritten as:
\ba
 h^{(0)}&=&-\frac{1}{2}(\partial_1^2 + \partial_2^2)
+\frac{1}{2} \gamma(\gamma+1) \{ \frac{1}{y_1^2}+
   \frac{1}{(\frac{1}{2}y_1+  \frac{\sqrt{3}}{2}y_2)^2}+
   \frac{1}{(-\frac{1}{2}y_1+ \frac{\sqrt{3}}{2}y_2)^2} \} +\nonumber\\
&+& 72\alpha^2 (y_1^2 + y_2^2), \label{scalar}
\ea
or, equivalently:
\be
h^{(0)}= 
A_1^+A_1^-+A_2^+A_2^-,
\label{factor}
\ee
where $(A_1^-, \, A_2^-)$ are the components of the vector operator 
\be
q^+_{(0,1)}\equiv (A_1^-, \, A_2^-),
\ee
which can be expressed in terms of the superpotential as:
\ba
A_1^- = (A_1^+)^{\dagger}
&\equiv & \frac{1}{\sqrt{2}} (\partial_1 + 
\partial_1 w(y_1,y_2) ) \equiv \nonumber\\
&\equiv & \frac{1}{\sqrt{2}} \biggl(\partial_1 +
12\alpha y_1 +
\frac{\gamma}{2}
 \biggl[ \frac{2}{y_1}+
   \frac{1}{\frac{1}{2}y_1+  \frac{\sqrt{3}}{2}y_2}-
   \frac{1}{-\frac{1}{2}y_1+ \frac{\sqrt{3}}{2}y_2} \biggr]\biggr); \nonumber\\
A_2^- = (A_2^+)^{\dagger}
&\equiv & \frac{1}{\sqrt{2}} (\partial_2 + 
\partial_2 w(y_1,y_2) ) \equiv \nonumber\\
&\equiv & \frac{1}{\sqrt{2}} \biggl(\partial_2 +
12\alpha y_2 +
\frac{\sqrt{3}\gamma}{2}
\biggl[ \frac{1}{\frac{1}{2}y_1+  \frac{\sqrt{3}}{2}y_2} +
   \frac{1}{-\frac{1}{2}y_1+ \frac{\sqrt{3}}{2}y_2} \biggr]\biggr) . 
\nonumber
%\label{charge}
\ea

The Hamiltonian $h^{(0)}$ is not symmetric
under the exchange of variables $y_1,y_2,$
however its wave functions can be obtained from
the well known wave functions of Calogero 
Hamiltonian \cite{calogero},
which are symmetric under the permutations of $x_i\,\, (i=1,2,3)$.

According to Section 2, the Hamiltonian $h^{(0)}$ generates a chain 
which includes a second scalar Hamiltonian defined (apart from a constant)
by:
\ba
h^{(2)} &=& B_1^+B_1^- + B_2^+B_2^- 
=
 -\frac{1}{2}(\partial^2_1+\partial_2^2) +\nonumber\\
&+&\frac{1}{2} \gamma(\gamma-1) \{ \frac{1}{y_1^2}+
   \frac{1}{(\frac{1}{2}y_1+  \frac{\sqrt{3}}{2}y_2)^2}+
   \frac{1}{(-\frac{1}{2}y_1+ \frac{\sqrt{3}}{2}y_2)^2} \} +
72\alpha^2 (y_1^2 + y_2^2), \label{scalar2}
\ea
where we have introduced the operators 
$B_l^{\pm}\equiv \epsilon_{lk}A_k^{\mp};\quad \epsilon_{12}
=-\epsilon_{21}=1; \epsilon_{11}=\epsilon_{22}=0.$

Also included in the chain is the $2\times 2$ matrix Hamiltonian:
$$
h^{(1)}_{ik}=
A_i^-A_k^+ + B_i^-B_k^+
$$
\ba
h^{(1)}&=&
-\frac{1}{2}(\partial_1^2+\partial_2^2)+
\frac{1}{2}[(\partial_lw)^2- \partial_l^2w]
+\frac{1}{2} \left( \matrix{
\partial_1^2w &  \partial_1\partial_2w  \cr
\partial_1\partial_2w& \partial_2^2w 
} \right)=          \nonumber \\
&=& -\frac{1}{2}(\partial_1^2+\partial_2^2)+
72\alpha^2(y_1^2+y_2^2)+36\alpha\gamma
+ \frac{\gamma^2 -\sigma_3\gamma }{y_1^2}+\nonumber\\
&+&\frac{\gamma^2-\frac{1}{2}\gamma\sigma_3-\frac{\sqrt{3}}{2}\gamma\sigma_1}
{(\frac{1}{2}y_1+  \frac{\sqrt{3}}{2}y_2)^2}
+\frac{\gamma^2-\frac{1}{2}\gamma\sigma_3+\frac{\sqrt{3}}{2}\gamma\sigma_1}
   {(-\frac{1}{2}y_1+  \frac{\sqrt{3}}{2}y_2)^2}
\label{matrix}
\ea
where $\sigma_i$ are Pauli matrices.

The above Hamiltonians $h^{(0)}$ and $h^{(2)}$ are intertwined
with $h^{(1)}:$
\ba
h^{(0)}A_l^+ &=& A_k^+ h^{(1)}_{kl};
\qquad A_l^-h^{(0)} = h^{(1)}_{lk}A_k^-; \nonumber\\
h^{(2)}B_l^+ &=& B_k^+ h^{(1)}_{kl};\qquad
B_l^-h^{(2)} = h^{(1)}_{lk}B_k^-; \qquad
l,k=1,2,
\label{intertw}
\ea
This chain of Hamiltonians $h^{(0)}, h^{(1)}_{lk}, h^{(2)}$ determines
the superHamiltonian as a Schr\"odinger-like operator with  $4\times 4$ 
matrix potential of block-diagonal form. 
Intertwining relations (\ref{intertw}) lead to interrelations between
spectra and eigenfunctions of the chain Hamiltonians. Apart from 
possible zero modes
of $A_l^{\pm},\, B_l^{\pm},$ the spectrum of $2\times 2$ matrix 
Hamiltonian $h^{(1)}$ is formed by two parts, coinciding with the spectra
of the scalar Hamiltonians $h^{(0)}$ and $h^{(2)},$ correspondingly.
Their eigenfunctions\footnote{$\Psi^{(1)}_l(E^{(0)})$ are the components
$(l=1,2)$ of two-component vector eigenfunctions of the 
matrix Hamiltonian $h^{(1)}.$ } 
are connected by the intertwining operators:
\ba
\Psi^{(1)}_k(E^{(0)})\sim A_k^-\Psi^{(0)}(E^{(0)}); \qquad 
\Psi^{(1)}_k(E^{(2)})\sim B_k^-\Psi^{(2)}(E^{(2)});\nonumber\\ 
\Psi^{(0)}(E^{(0)})\sim A_k^+\Psi^{(1)}_k(E^{(0)}); \qquad
\Psi^{(2)}(E^{(2)})\sim B_k^+\Psi^{(1)}_k(E^{(2)}). 
\label{psi}
\ea
Thus all (up to zero modes of the
$A_l^{\pm},\, B_l^{\pm}$) eigenvectors of the 
matrix Hamiltonian $h^{(1)}$ are expressed
in terms of the Calogero wave functions.

In summary, we have used the framework of SUSY QM in order to derive 
an exactly solvable $2\times 2$ matrix model, the spectrum of which
is divided into two parts, each one coinciding with the 
spectrum of a scalar Calogero Hamiltonian. 
The reason why the spectrum of the matrix model (\ref{matrix}) is still
completely discrete can be found in the dominance of the confining scalar 
interaction over the coupling of internal degrees of freedom which is
asymptotically decreasing.
This matrix problem in a non-trivial way is related to a system of
independent harmonic oscillators \cite{gurappa}, but is not
diagonalizable by standard transformations like rotations.

\section*{\large\bf 4.\quad Time-dependent exactly solvable $3$-body matrix 
problems}
\hspace*{3ex}
In this Section we will achieve the goal of obtaining scalar and matrix
time-dependent exactly(quasi-exactly) solvable models and  
invariant operators. 
We start from a general time-dependent intertwining relations
which connect two time-dependent Schr\"odinger equations (TDSE)
\footnote{Both Hamiltonians are assumed to be hermitian.}, one
of them with a time-independent exactly-solvable Hamiltonian. 
If $H(\vec{y})$ is an exactly solvable Hamiltonian and
$H(\vec{y}) \psi_n(\vec{y})=E_n\psi_n(\vec{y}),$ the intertwining 
relation with a known operator $U(\vec{y},t):$
\ba
(i\partial_t - \tilde{H}(\vec{y},t)) U(\vec{y},t) = U(\vec{y},t) 
(i\partial_t - H(\vec{y})) \label{t_intertw}
\ea
leads to an exactly solvable time-dependent problem.
All the solutions of 
$$(i\partial_t - \tilde{H}(\vec{y},t))
\tilde{\Psi} (\vec{y},t) = 0 $$ 
can be written as $U(\vec{y},t) \Psi (\vec{y},t),$ where $\Psi (\vec{y},t) 
= \sum_{n=0}^{\infty} c_ne^{-iE_nt}\psi_n(\vec{y})$
is a generic time-dependent solution of equation 
$ (i\partial_t - H(\vec{y}))\Psi (\vec{y},t) = 0.$ 
 
For the one-dimensional case intertwining relations (\ref{t_intertw})
were investigated in \cite{tdse} for differential operators $U(y,t)$
of first and second order in derivatives. While in the one-dimensional
problem a wide class of solutions was found, a straightforward extension
to the two-dimensional case does not appear to be obvious. In this case 
it is more effective to study operators $U(\vec{y},t)$ which can be written
as products of two unitary
pseudo-differential (of infinite order in derivatives) operators of the form
\cite{china}:
\ba
U(\vec{y},t) \equiv \exp\{ia(t)\sum_{i}y_i^2\}\cdot\exp\{b(t)
\sum_{i}(y_i\partial_i+\partial_iy_i)\},  \label{unitary}
\ea
where $a(t),\,b(t)$- are arbitrary external time-dependent real functions.
These operators have no zero modes. The 
intertwining relation (\ref{t_intertw}) leads to:
\ba
\tilde{H}(\vec{y},t)) = U(\vec{y},t)H(\vec{y})U^{-1}(\vec{y},t) +
i (\frac{\partial U(\vec{y},t)}{\partial t})U^{-1}(\vec{y},t) .
\label{connection}
\ea

In the supersymmetric framework (Sections 2 and 3) for each Hamiltonian of the 
chain one can choose the real valued coefficient functions 
$a^{(M)}(t),$ $b^{(M)}(t)$ independently for the different values of $M.$  
Under these unitary 
transformations $U^{(M)}$ the Jacobi canonical variables transform as:
\ba
y_i&\rightarrow & U^{(M)}y_i(U^{(M)})^{-1} = 
y_i\cdot\exp\{2b^{(M)}(t)\};\nonumber\\
p_i\equiv -i\partial_i 
&\rightarrow &  U^{(M)}p_i(U^{(M)})^{-1} = 
(p_i-2a^{(M)}(t)y_i)\cdot\exp\{-2b^{(M)}(t)\},
\nonumber
%\label{coordinates}
\ea
and the so called gauge term in (\ref{connection}) reads:
\be
i (\frac{\partial U^{(M)}(\vec{y},t)}{\partial t})(U^{(M)})^{-1}(\vec{y},t) =
\bigl(4a^{(M)}(t)\dot{b}^{(M)}(t) - \dot{a}^{(M)}(t)\bigr)\sum_{i}y_i^2 -
\dot{b}^{(M)}(t)\sum_{i}(y_ip_i+p_iy_i).
\label{gauge}
\ee

After setting up the general framework of time-dependent intertwining 
of TDSE, we apply it to  
the Hamiltonians $h^{(M)}$ of the Calogero superchain
of Section 3. 
In particular, we identify $H(\vec{y})$ with the elements $h^{(M)}$
of the $3$-body Calogero chain $M=0,1,2$ and generate a time-dependent
chain. In general, the time-dependent Hamiltonians 
acquire new terms linear in momenta and  time dependent 
coefficients in all terms:
\ba
\tilde{h}^{(0)}(\vec{y},t) &=&\frac{1}{2} e^{-4b^{(0)}(t)}\sum_{i=1,2}p_i^2 -
\biggl( a^{(0)}(t)e^{-4b^{(0)}(t)}+\dot{b}^{(0)}(t) \biggr)
\sum_{i=1,2}(y_ip_i+p_iy_i)+\nonumber\\ &+& 
\biggl( 2(a^{(0)}(t))^2e^{-4b^{(0)}(t)}
+72\alpha^2e^{4b^{(0)}(t)} + 4a^{(0)}(t)\dot{b}^{(0)}(t) - 
\dot{a}^{(0)}(t) \biggr)\sum_{i=1,2}y_i^2+ \nonumber\\ 
&+&\frac{1}{2} e^{-4b^{(0)}(t)}
\gamma(\gamma +1) \biggl[ \frac{1}{y_1^2}+
   \frac{1}{(\frac{1}{2}y_1+  \frac{\sqrt{3}}{2}y_2)^2}+
   \frac{1}{(-\frac{1}{2}y_1+ \frac{\sqrt{3}}{2}y_2)^2} \biggr]. 
\label{tscalar}
\ea
%{\it Write and investigate the condition that} 
%$\tilde{H}(x,t=0)\equiv H^(x)$ The conditions are:$b^{(0)}(0)=0\,
%\dot{a}^{(0)}(0)=-2(a^{(0)})^2;/,\dot{b}^{(0)}(0)=-a^{(0)}(0) $
The second scalar Hamiltonian $\tilde{h}^{(2)}(\vec{y},t)$ 
results from (\ref{scalar2}) with a similar construction.

The matrix Hamiltonian of the chain has the form:
\ba
\tilde{h}^{(1)}(\vec{y},t)&=&\frac{1}{2} e^{-4b^{(1)}(t)}\sum_{i=1,2}p_i^2 -
\biggl( a^{(1)}(t)e^{-4b^{(1)}(t)}+\dot{b}^{(1)}(t) \biggr)
\sum_{i=1,2}(y_ip_i+p_iy_i) +\nonumber\\&+& 
\biggl(2(a^{(1)}(t))^2e^{-4b^{(1)}(t)}
+72\alpha^2e^{4b^{(1)}(t)} + 4a^{(1)}(t)\dot{b}^{(1)}(t) - 
\dot{a}^{(1)}(t) \biggr)\sum_{i=1,2}y_i^2 + 36\alpha\gamma +
\nonumber\\&+& 
e^{-4b^{(1)}(t)}
\biggl[\frac{\gamma^2 -\sigma_3\gamma }{y_1^2}+
\frac{\gamma^2-\frac{1}{2}\gamma\sigma_3-\frac{\sqrt{3}}{2}\gamma\sigma_1}
{(\frac{1}{2}y_1+  \frac{\sqrt{3}}{2}y_2)^2}
+\frac{\gamma^2-\frac{1}{2}\gamma\sigma_3+\frac{\sqrt{3}}{2}\gamma\sigma_1}
   {(-\frac{1}{2}y_1+  \frac{\sqrt{3}}{2}y_2)^2}\biggr]. \label{tmatrix}
\ea

The $t$-dependence of the kinetic term can be interpreted as
a $t$-dependent mass \cite{nieto}.
Linear terms in momenta are known to describe for example the coupling 
of charged
particles with gauge potentials and therefore have not to be discarded
a priori. However the terms linearly dependent on momenta
drop out for a particular relation:
\be 
a(t)=-\dot{b}(t)\exp(4b(t)).
\label{ab}
\ee

We remark that, in the case of 
the factorization of $\tilde{h}^{(M)}(\vec{y},t) = \eta (t)
h^{(M)}(\vec{y}),$ TDSE reduces effectively to a quasi-stationary
problem, because by a suitable reparametrization
of time $t\rightarrow \tau\equiv \int \eta(t)dt$ the problem becomes 
stationary.
The corresponding constraint leads for $M=0,1,2$ again to (\ref{ab})
and to a nonlinear differential equation for $ b(t) :$
\be
\dot{}b\dot{}(t)+6\dot{b}^2(t)+72\alpha^2(e^{-8b(t)}-1)=0.
\nonumber
%\label{kamke}
\ee 
The general solution of this equation involves elliptic integrals in the
relation between $t$ and $b.$ The function $\eta (t)$ becomes 
$\eta (t) = \exp(-4b(t)).$

The construction of invariant operators $R$, which satisfy the equation:
\be
\frac{\partial R}{\partial t}+ i[ \tilde{H}(\vec{y},t), R ]=0 
\nonumber
\ee
is an important aspect
of the investigation of time-dependent systems \cite{kaushal}.
In our framework from the intertwining relation (\ref{t_intertw}) 
the invariant operator exists and can be expressed in terms of $h^{(M)}$
and $U^{(M)}$:
\ba
R^{(M)}(t)&\equiv & U^{(M)}(\vec{y},t)h^{(M)}(\vec{y}) 
(U^{(M)})^{-1}(\vec{y},t)= \label{invariant}\\
&=& \tilde{h}^{(M)}(\vec{y},t) - 
i(\frac{\partial U^{(M)}(\vec{y},t)}{\partial t})
(U^{(M)})^{-1}(\vec{y},t)\nonumber
\ea
where the last term is usually referred as gauge term. 

The invariant operator is hermitian because the intertwining
operator $U(\vec{y},t)$ is unitary. From the Eqs.(\ref{tscalar}),
(\ref{tmatrix}) and (\ref{gauge}) it is straightforward to obtain
the explicit expression for the chain of invariants of this model.
In particular, one can notice that $R^{(M)}$ have still the structure 
similar to the Calogero Hamiltonians (\ref{tscalar}), (\ref{tmatrix})
though some terms are missing. 

In general, one can argue from the similarity (\ref{invariant}) 
that the spectrum of $R^{(M)}$ is the same
as the spectrum of $h^{(M)}$ and therefore time-independent \cite{lewis}.
The operators $R^{(M)}$ provide an 
additional exactly solvable (matrix and scalar) 
models with explicit time-dependent potentials but with time-independent
spectra. Their
eigenfunctions depend parametrically on time via $U^{(M)}(\vec{y},t)$
applied to the stationary eigenfunctions of $h^{(M)}.$  
Let us remark that invariant operators $R(t)$ admit a quasi-factorization
like (\ref{factor}) in Section 3 with suitable (transformed by 
$U(\vec{y},t))$ components of supercharge, 
but $\tilde{h}(\vec{y}, t)$ do not because of the gauge term.

\section*{\large\bf 5.\quad Conclusions}
\hspace*{2ex} 

Given for granted the usefulness of exactly (and quasi-exactly) 
solvable models we would like to point out that our contribution
has been to construct explicitly few models of such a kind with a discrete
spectrum:
among them the exactly solvable $3$-particle (matrix
and scalar) non-stationary Calogero models and quasi-exactly-solvable
$N$-particle matrix stationary models.    
An extension of the method of Section 4 to the $N$-body Calogero model
described in Section 3 leads to time-dependent quasi-exactly-solvable
matrix models. 
Since it is not usual to find exactly solvable or quasi-exactly solvable
time-dependent problems, specially in a context of the many-body
systems, our results support the program to investigate
further time-dependent generalizations of stationary solvable models,
like mentioned in the Introduction \cite{sasaki}, 
\cite{q}, \cite{znojil}, \cite{multi}
and quasi-exactly solvable matrix models \cite{brihaye}. 
In particular, one can focus attention on the dynamical algebras
of these models \cite{andric} to construct Ermakov-Lewis invariant
operators (\cite{kaushal} and references therein).
A less straightforward task will be to modify the model 
in such a way as to allow for coexistence \cite{acdi} of a continuum and a 
discrete spectrum describing scattering and bound states. 

\section*{\large\bf\quad Acknowledgments.}
\hspace*{3ex}
One of the authors (M.I.) thanks INFN and University of Bologna 
for warm hospitality and A.Neelov for useful discussions.
This work was partially supported by RFBR (grant 99-01-00736).

\end{document}